\documentclass[prd, aps, nofootinbib, preprintnumbers,
superscriptaddress,twocolumn, 10pt, floatfix]{revtex4-1}
\usepackage{graphics,graphicx,rotating}
\usepackage[usenames]{color}
\usepackage[percent]{overpic}
\usepackage{mathrsfs,amsmath,amssymb}
\usepackage{times}
\usepackage{mathptmx}
\usepackage{bm,multirow}

\newcommand{\be}{\begin{equation}}
\newcommand{\ee}{\end{equation}}
\newcommand{\beq}{\begin{equation}}
\newcommand{\eeq}{\end{equation}}
\newcommand{\ber}{\begin{eqnarray}}
\newcommand{\eer}{\end{eqnarray}}
\newcommand{\bea}{\begin{eqnarray}}
\newcommand{\eea}{\end{eqnarray}}

\newcommand{\cardiff}{\affiliation{School of Physics and Astronomy, Cardiff
University, Cardiff, CF24 3AA, United Kingdom}}

\begin{document}

\title{Degeneracy between mass and spin in black-hole-binary waveforms}

\author{Emily Baird}            \cardiff
\author{Stephen Fairhurst} \cardiff
\author{Mark Hannam}        \cardiff
\author{Patricia Murphy}     \cardiff

\date{\today}

\begin{abstract}

We explore the degeneracy between mass and spin in gravitational
waveforms emitted by black-hole binary coalescences.  We focus on 
spin-aligned waveforms and obtain our results using phenomenological 
models that were tuned to numerical-relativity simulations.  A degeneracy
is known for low-mass binaries (particularly neutron-star binaries),
where gravitational-wave detectors are sensitive to only the inspiral
phase, and the waveform can be modelled by post-Newtonian theory.  Here,
we consider black-hole binaries, where detectors will also be sensitive to 
the merger and ringdown, and demonstrate that the
degeneracy persists across a broad mass range.  At low masses, the
degeneracy is between mass ratio and total spin, with chirp mass
accurately determined.  At higher masses, the degeneracy persists but is not
so clearly characterised by constant chirp mass as the merger and ringdown become 
more significant. We consider
the importance of this degeneracy both for performing searches
(including searches where only non-spinning templates are used) and in
parameter extraction from observed systems.  We compare observational
capabilities between the early ($\sim$2015) and final (2018 onwards)
 versions of the Advanced LIGO detector.  

\end{abstract}

\maketitle

\section{Introduction}

The advanced LIGO and Virgo detectors are likely to allow us to observe
numerous black-hole-binary coalescences in the coming 
years~\cite{Abadie:2010cf, Dominik:2012vs}. 
While the detectors are being installed, the challenge is to devise search and
source extraction methods that will identify all black-hole-binary
signals in the data whilst also allowing for the most accurate estimation of the
source parameters possible. 

Our understanding of gravitational waveforms emitted by black-hole 
mergers has improved dramatcially over the past five years, with the aid of
a large number of numerical simulations, from which 
several models of the waveforms have been 
constructed (see e.g., Ref~\cite{Ohme:2011rm} for an overview).  
A large fraction of the simulations have focussed on
non-precessing systems (where the black holes are either non-spinning or have spins
aligned with the orbital angular momentum); see, for example, 
Refs.~\cite{Hannam:2009rd} and \cite{Ajith:2012tt}.  Other than for high
black-hole spins (greater than $a\sim0.8$), this space has been rather well
covered for comparable-mass binaries, and several waveform models are 
available~\cite{Ajith:2009bn,Santamaria:2010yb,Pan:2009wj}. 
The models characterize
waveforms based on the masses of the two components and, in the
case of the models we will consider in this paper, the total
spin of the system.  We will make use of these waveform models to
investigate degeneracies in the waveforms.  Since we are
only considering a three-dimensional subspace of the full 
eight-dimensional space of binary masses and spins, many effects, most notably
precession, cannot be probed.  However, it is interesting to consider
this smaller space because it is very likely that any degeneracies will
persist in the full parameter space.  Furthermore, recent results have
shown that the waveform for a precessing binary can be factorized into
precessional effects and a non-precessional part that is well modelled
by the three parameters we consider here~\cite{Schmidt:2012rh}.

In post-Newtonian (PN) theory, there is a well known degeneracy between 
the black holes' mass ratio and the total spin, which arises at 1.5PN order,
while a combination of the total 
mass and mass ratio (the ``chirp mass'') remains
relatively well constrained~\cite{Cutler:1994ys,Poisson:1995ef}. 
Here, our focus is on higher-mass waveforms that include a merger and
ringdown.  We demonstrate that the degeneracy persists in the full
waveforms and also up to higher masses where only the later part of the
inspiral and the merger/ringdown are in the detector's sensitivity band.
However, for the high-mass binaries, the degeneracy in mass and spin is
not so clearly characterized by the chirp mass.

A degeneracy in the emitted gravitational waveform across the parameter space
has numerous consequences for gravitational wave (GW) searches, both good and
bad.  The positive effect is that a degeneracy in the parameter space will reduce 
the volume that needs to
be searched, thereby reducing the computational cost and trials factor
associated with the search.  The majority of searches of LIGO and Virgo data
have made use of non-spinning components in the template waveforms (see
e.g., Refs.~\cite{Collaboration:S6CBClowmass, Collaboration:S5HighMass}).  A
degeneracy between mass and spin means that the template waveforms have covered
a larger fraction of the parameter space than might be naively expected.
Furthermore, it has recently been argued that a two-dimensional bank of
templates is sufficient to cover the space of spinning neutron-star
binaries~\cite{Brown:2012qf}.  In this paper, we investigate the effect of
using non-spinning waveforms in a search for black-hole binaries with spins and
show that the mass-spin degeneracy renders the search more sensitive to spinning
systems than might be expected.  Nevertheless, a search using waveforms incorporating
spin effects would be a significant improvement and the degeneracy should make it 
computationally feasible.
 
On the other hand, a degeneracy has a negative effect on parameter estimation. 
To make the most of gravitational-wave observations, accurate extraction of the
physical parameters is of paramount importance.  There are detailed
multi-dimensional methods under development to accurately recover the
parameters \cite{Sluys:2008a, Sluys:2008b, Veitch:2010, Feroz:2009}.  However,
there's nothing that can be done about a real degeneracy in the emitted
waveforms --- there is no way of telling them apart.  We evaluate the effects
of mass-spin waveform degeneracy on our ability to accurately recover masses
and spins and discuss the astrophysical implications. In the process we introduce
a simple method to estimate the parameter-estimation confidence intervals. 

Throughout the paper, we will provide sample results using a number of
waveforms at different masses and mass ratios.  We also make use of a
number of noise curves for the advanced detectors, to illustrate how
this effect is likely to change as the detectors approach their final,
design sensitivity.  No attempt has been made to perform an exhaustive
study across the full parameter space, and this is left as a future
project.

The layout of the paper is as follows: in Sec.~\ref{sec:model} we
outline the waveform models that we use, and discuss some of the
assumptions that went into producing them and their range of
applicability.  We also discuss the interpretation of results in terms
of mismatches between waveforms and introduce the detector noise curves
used in our studies.  In Sec.~\ref{sec:spin_eta} we focus on the
degeneracy between mass ratio and spin for low-mass binaries, giving the
post-Newtonian argument for this degeneracy and illustrating the
degeneracy for a number of cases and different detector sensitivities,
and discuss implications for searches.  Then, in Sec.~\ref{sec:high_mass} 
we extend the results to higher mass binaries and
show that the degeneracy persists, although in a different form.  In
Sec.~\ref{sec:parameters} we discuss implications for the accurate
estimation of parameters and astrophysical inference.


\section{Model and methods}
\label{sec:model}

\subsection{The phenomenological waveform models} 
\label{sec:phenom}

We describe the GW signal from black-hole binaries with non-precessing spins 
(i.e., the spins are aligned or anti-aligned with the binary's orbital angular momentum) 
using the phenomenological models presented in~Refs.~\cite{Ajith:2009bn}
and \cite{Santamaria:2010yb}. For consistency with the labelling used within the 
LIGO-Virgo Collaboration \cite{LAL} 
we refer to these models respectively as 
``PhenomB'' and ``PhenomC''. (``PhenomA'' refers to an earlier model of non-spinning
binaries~\cite{Ajith:2007qp,Ajith:2007kx,Ajith:2007xh}.) 
In both models the waveforms are parametrized by 
their total mass $M = m_1 + m_2$, mass ratio $\eta = m_1 m_2 / M^2$, 
and an effective total spin parameter, 
\begin{eqnarray}
\chi & = & \frac{1}{M} (m_1 \chi_1 + m_2 \chi_2) \nonumber \\
     & = & \chi_{s} + \delta \chi_{a},
\end{eqnarray} 
where $\chi_i = S_i / m_i^2$ for each black hole with
angular momentum $S_i$, $\delta = (m_1 - m_2)/M$, and the symmetric and anti-symmetric 
combinations of the
spins are $\chi_s = (\chi_1 + \chi_2)/2$ and $\chi_a = (\chi_1 - \chi_2)/2$.
The phenomenological models incorporate a PN description of the inspiral, while
the merger and ringdown regimes are tuned using the results of numerical
simulations.  

Both models have the same basic structure. The waveform is represented in the Fourier
domain as $h(f) = A(f) e^{i \Psi(f)}$.  The amplitude $A(f)$ and phase $\Psi(f)$ are modelled
separately, using input from PN theory (inspiral), the observed properties of NR waveforms 
(plunge-merger), or results from perturbation theory (ringdown).  
The models are all power series in the frequency $f$, and the coefficients in the model are 
written as polynomials in the two physical parameters $\eta$ and $\chi$ (the total mass is an overall 
scale factor), and it is the coefficients of these polynomials that are then calibrated to hybrids of
PN and NR waveforms.  For both models, the amplitude is constructed in a similar manner, with 
expressions for each of the inspiral, plunge-merger, and ringdown portions of the waveform
modelled independently.  In PhenomB~\cite{Ajith:2009bn} the three parts are connected
as piecewise functions, and in PhenomC~\cite{Santamaria:2010yb} smooth $\tanh$-function 
interpolation is used.  To obtain an expression for the phasing, PhenomB uses a single series
expansion, matching the coefficients beyond leading order to hybrid waveforms, as well as 
results from the test mass ($\eta \rightarrow 0$) limit.  PhenomC, on the other hand, 
uses the complete TaylorF2 PN inspiral phasing, and only
the late inspiral/merger phase is fitted in a narrow frequency range to numerical 
simulations, while the ringdown waveform is obtained from analytically derived quasi-normal mode
expressions for the frequency and attached continuously to the merger phase.
There are 54 free parameters in PhenomB, and 45 in PhenomC, although the final 
models are both functions of only $\{M,\eta,\chi\}$. 
There are three notable differences between the two models: (1) the PhenomB PN-NR hybrids
are produced in the time domain, using TaylorT1 for the PN part, and the 
PhenomC hybrids are produced in the frequency domain, using TaylorF2 for the 
PN part~\cite{Ajith:2007jx};  (2) PhenomB incorporates 
information about the test-mass limit;  (3) in PhenomC the phase evolution during inspiral 
incorporates PN calculations up to 3.5PN order (although the spin terms are complete only up 
to 2.5PN), while in PhenomB only the leading-order PN inspiral term is fixed, and the 
remaining terms up to 3.5PN order are tuned to the hybrid waveforms. 

There is good qualitative agreement between the two models~\cite{Santamaria:2010yb}, although no
detailed quantitative comparison has yet been performed.  We choose to work with PhenomB for 
generating the results presented in this paper. We have also cross-checked
some of the calculations against PhenomC, and we comment further on this in 
Sec.~\ref{sec:high_mass}.

\subsection{Detectors and noise curves} 
\label{sec:detectors}

\begin{figure}
\begin{center}
  \includegraphics[width=0.5\textwidth]{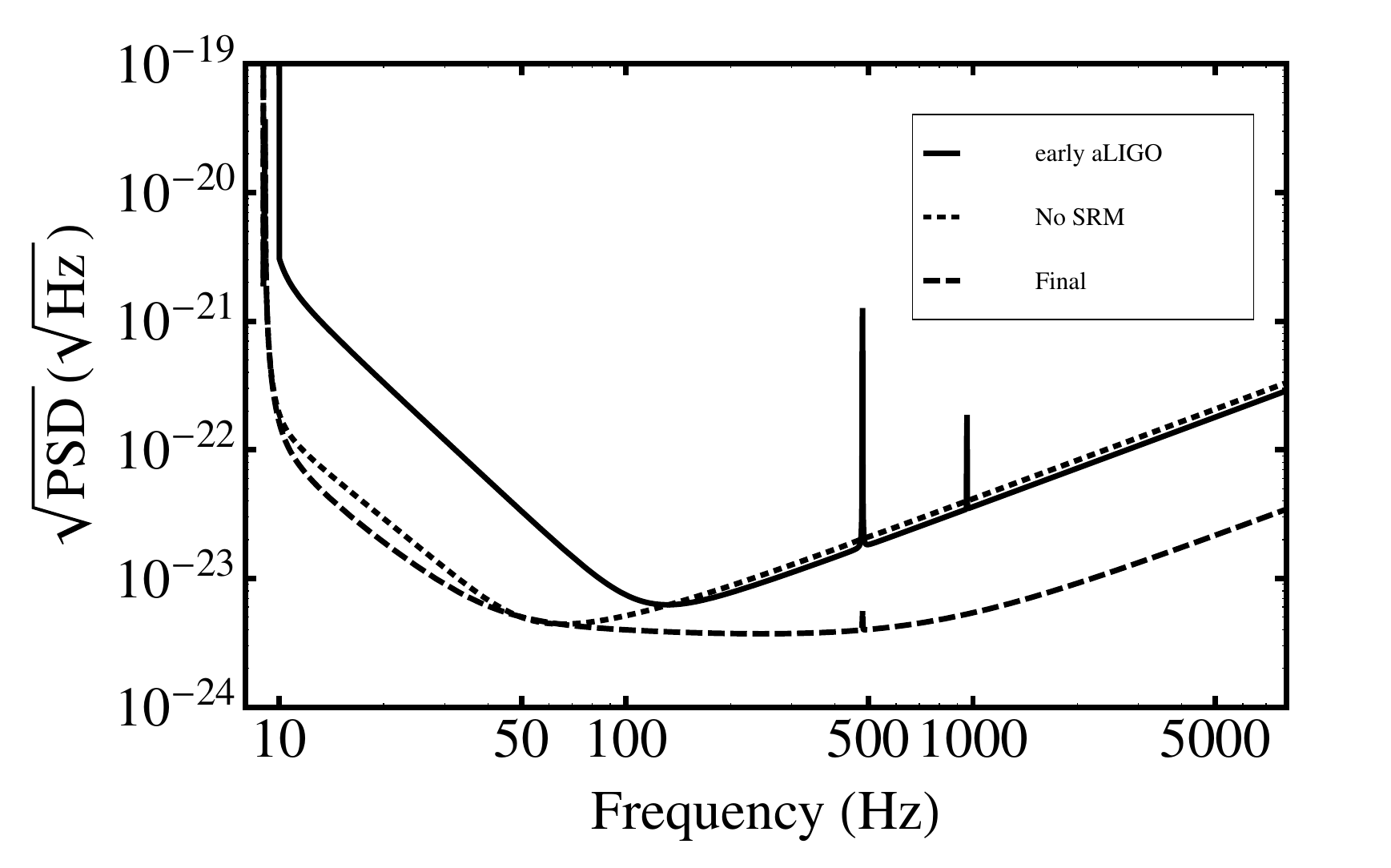}
  \caption{
  \label{fig:NoiseCurves} 
  Noise curves for the advanced LIGO detector configurations that we consider 
  in this paper: ``early'', ``no-SRM'' and ``final'', which corresponds to the 
  zero-detuned, high-power configuration. See text 
  for more details. 
  }
\end{center}
\end{figure}

In this paper we compare spinning- and non-spinning-binary signals with
reference to the expected sensitivity of the Advanced LIGO detector
(aLIGO)~\cite{Abbott:2007kv,Shoemaker:aLIGO,2010CQGra..27h4006H}.  The
sensitivity curves we use are shown in Fig.~\ref{fig:NoiseCurves}.  During the
early science runs, expected around 2015, the advanced LIGO detector is unlikely to be at
its full design sensitivity.  Consequently, we use the ``early aLIGO'' noise
curve~\cite{G1000176} to give results indicative of what may be achieved in the
early runs.   At its optimum sensitivity several years later, the anticipated
sensitivity is given by the ``zero-detuned high-power'' noise
curve~\cite{T0900288}. In this paper we will take that as the ``final'' design 
sensitivity of the detector.  Over the parameter space of binaries that we
study in this paper, the early aLIGO curve represents a sensitivity of
roughly five times greater in the signal-to-noise ratio (SNR) than in the
initial LIGO detectors during their final S6 science run, which corresponds to
an increase in potential sources of two orders of magnitude.  The final curve
represents a further factor of three improvement in SNR, or about 30 times as
many potential sources over early  aLIGO. 
Finally, we also consider the ``no signal recycling''
(no-SRM) configuration of the detector, that could be achieved by the aLIGO
detector operating without a signal recycling cavity. 
This has comparable 
low-frequency sensitivity to the final configuration, but significantly worse
sensitivity at high frequencies.  
Although it is unlikely to be an observational mode, the non-signal-recycled curve 
offers a means to compare the effects of low
and high frequency sensitivity upon our results.

\subsection{Waveform Mismatches} 
\label{sec:matches}

We use the standard inner product between two waveforms, $h_1(f)$ and $h_2(f)$
with respect to the power spectral density $S_n(f)$ of a
detector~\cite{Cutler94}, 
\begin{equation} (h_1 | h_2 ) = 4 \, {\rm Re} \int_{0}^{\infty} \frac{h_1(f) h_2^*(f)}{S_n(f)} df.  
\end{equation}
The match between two
waveforms is defined as their normalized inner product, maximized over time and
phase shifts of the waveform, 
\begin{equation} 
M(h_1,h_2) =  \max_{\Delta t, \Delta \phi} \frac{(h_2 | h_2)}{ | h_1 | | h_2| }.  
\end{equation}

Alternatively, we can consider the situation of a true waveform produced by a
physical source, $h_T$, and a model waveform $h_M$ that we will use to search
for the real signal in detector data. If we normalize both waveforms, then we
can split the model waveform into a component parallel to the true waveform,
plus a component that is orthogonal to the true waveform, in the sense of our
inner product. In other words, we have 
\begin{equation} 
\hat{h}_M = \sqrt{1 - x^2} \hat{h}_T + x \hat{h}_E, 
\end{equation} 
where $\hat{h}_E$ is the normalized
``error'' waveform that satisfies $(h_T | h_E) = 0$.  In this form, the match is
$M(h_T,h_M) = \sqrt{1 - x^2} \approx 1 - x^2/2$.  Thus, the match is directly 
related to the (relative) amplitude of the ``error'' waveform.

In a GW search, a template bank of model waveforms is constructed
\cite{Babak:2006dw} such that the match between every point in the waveform
parameter space and the nearest template in the bank is at least 0.97. Assuming
that the model waveforms are physically correct, such a template bank ensures
that we will lose no more than 10\% of signals in our search.  (A match of 0.97
means that the sensitivity range of the detector is only 97\% of its optimum,
and the detector is therefore sensitive to only $0.97^3 \approx 0.9$ of its
optimum volume, and so we lose about 10\% of signals.)

If the physical waveforms do not agree exactly with the model waveforms, this
will lead to an additional loss in match between a signal and the best matched
template, and consequently a reduction in the number of signals observed above
threshold.  In addition, to counter non-stationary detector noise, a
number of signal consistency tests are included into analyses to distinguish
signals from non-stationarities or ``glitches'' in the data \cite{Babak:2012zx}.
These tests are used to either remove completely any transients that do not
match the templates or else to down-weight their significance.  Since searches
performed to date have made use of non-spinning waveform families these
thresholds have been set relatively loosely (and tested with spinning signals)
to ensure they are not removing signals.  For the high matches between signal and
template we consider in this paper, the effect of signal consistency tests will be
minimal, and we will not consider it further.

In the next sections we will identify the regions of the non-spinning waveform
parameter space that provide a match of greater than 0.97 with the chosen
waveform (which will usually incorporate spin).  In doing so, we densely sample
the non-spinning waveforms to identify all points with a match above 0.97.  When
performing a search, there will then be two contributions to the mismatch
between the signal and the best matched template: one due to the difference
between the waveform and search space and a second arising from the discrete
sampling of the template space.  The match between the signal and the closest
template is guaranteed to be above 0.94 as the mismatches add linearly in this
case (see e.g.~\cite{Lindblom:2008ha} for details).  We are
therefore requiring that the potential loss of SNR due to a mismatch between the
model waveform and true waveform is no greater than the maximum possible loss
due to the discreteness of the template bank.

We will also investigate how the parameters of the best-match
\textit{non-spinning} waveform vary when the signal corresponds to a
non-precessing binary.  In Sec.~\ref{sec:parameters} we will see how the match
can be used to give an estimate of the parameter estimation accuracy.


\section{Degeneracy between $\eta$ and $\chi$ at low masses} 
\label{sec:spin_eta}

We first consider the degeneracy between the symmetric mass ratio $\eta$ and the effective total
spin $\chi$ of the binary. This effect is already well known in PN 
theory~\cite{Cutler:1994ys,Poisson:1995ef}, which we will discuss first, and
then we will look at inspiral-merger-ringdown (IMR) signals at three different values of the total mass, 
$M = \{20, 50, 100\} M_\odot$.

\subsection{Degeneracy in PN theory} 
\label{sec:PNtheory}

The phase evolution of a compact binary in PN theory has been calculated up to 3.5PN order in the
non-spinning terms, and up to 2.5PN in the spin effects; see Ref.~\cite{Ajith:2007jx} for a recent summary of 
PN treatments of the phase. Up to the leading order that 
includes spin, the phase for non-precessing binaries is given in the frequency domain by 
\begin{eqnarray*}
\Psi(f) & = & \frac{3}{128 \eta v^5} \left\{ 1 +  v^2 \left[ \frac{3715}{756} + \frac{55 \eta}{9} \right]  \right. \\
&& \ \ \ \ \ \ \ \ \ \ \ \ \left. - v^3 \left[ 16 \pi - \left( \frac{113}{3} - \frac{76 \eta}{3} \right) \chi_s - \frac{113 \delta}{3} \chi_a \right] \right\}
\label{eqn:PN}
\end{eqnarray*}
where $v = (\pi M f)^{1/3}$. If we define the chirp mass of the binary as
${\cal M} = M \eta^{3/5}$, then we see that the leading factor is proportional
to $1/({\cal M} \pi f)^{5/3}$, and the phase evolution is dominated by the
chirp mass.  This motivates the observation that in GW searches we expect to
measure the chirp mass with high accuracy, and we will see examples of this in
Sec.~\ref{sec:IMRlow}. At the next-to-leading order, the phase evolution
depends on the mass ratio $\eta$, and dependence on the spins enters at the 
following order.  We can absorb all of
the spin effects at this order into an effective spin term, $\chi_{\rm PN} =
\chi_s + \delta \chi_a - (76\eta/113) \chi_s$, and this is what is proposed in
the inspiral template family discussed in Ref.~\cite{Ajith:2011ec} 
(and this is proportional to the leading-order spin-orbit parameter $\beta$ used in
Refs.~\cite{Cutler:1994ys,Poisson:1995ef}). The
phenomenological models~\cite{Ajith:2009bn,Santamaria:2010yb} use the simpler
effective spin $\chi = \chi_s + \delta \chi_a$ to describe full inspiral,
merger and ringdown waveforms.

If we adopt for the moment the PN effective spin term, we can write the
deviation from the leading-order term as
\begin{equation}
\Delta \Psi(f)  = \frac{3}{128 \eta v^3} \left[ \frac{3715}{756} + \frac{55 \eta}{9} 
+ v \left( \frac{113 \chi_{\rm PN}}{3} - 16 \pi \right)  \right].
\label{eqn:PNdeg}
\end{equation}
Thus, it is possible to mimic the effect of spin by modifying the mass ratio
(while keeping the chirp mass constant).  It is clear, however, that the
required modification to $\eta$ will vary as $v$ changes over the course of the chirp, as
expected as this is only an approximate degeneracy.  The majority of the power
from a binary merger is accumulated between around 30 to 300 Hz, and over this
range $\nu$ changes by only a factor of two.  This explains why the
approximation is reasonable across the whole signal, and this leads to the
common claim that non-spinning templates can be used to detect GW signals from
spinning binaries, but there will be an offset in the measurement of the mass
ratio and the total mass. It may also be possible to exploit this
degeneracy to search for spinning binaries using a non-spinning model, but with
$\eta$ extended to unphysical values, $\eta > 0.25$, in order to cover more of
the spinning-binary parameter space; a similar idea has already been suggested
to extend inspiral-only searches beyond their expected region of
validity~\cite{Boyle:2009dg}.  It is the degeneracy between mass ratio and spin
that we will investigate here.

\subsection{Degeneracy in full IMR models} 
\label{sec:IMRlow}

We would now like to investigate whether the degeneracy from
post-Newtonian theory is present in full IMR waveform models.  

We consider a set of spinning signals and those non-spinning waveforms (i.e.
the phenomenological model with  $\chi = 0$) that provide a good match to the
signal.  The efficacy of a waveform model in a GW search can be estimated by
calculating the best match (fitting factor) between any member of the waveform
model, and the target signal.  Therefore, this will give an immediate indication
of the merits (or otherwise) of using a non-spinning waveform model to search
for mergers of spinning black holes.  We consider fitting factors above 0.97 to
be adequate as discussed previously.

\begin{figure}
\begin{center}
  \includegraphics[width=0.5\textwidth]{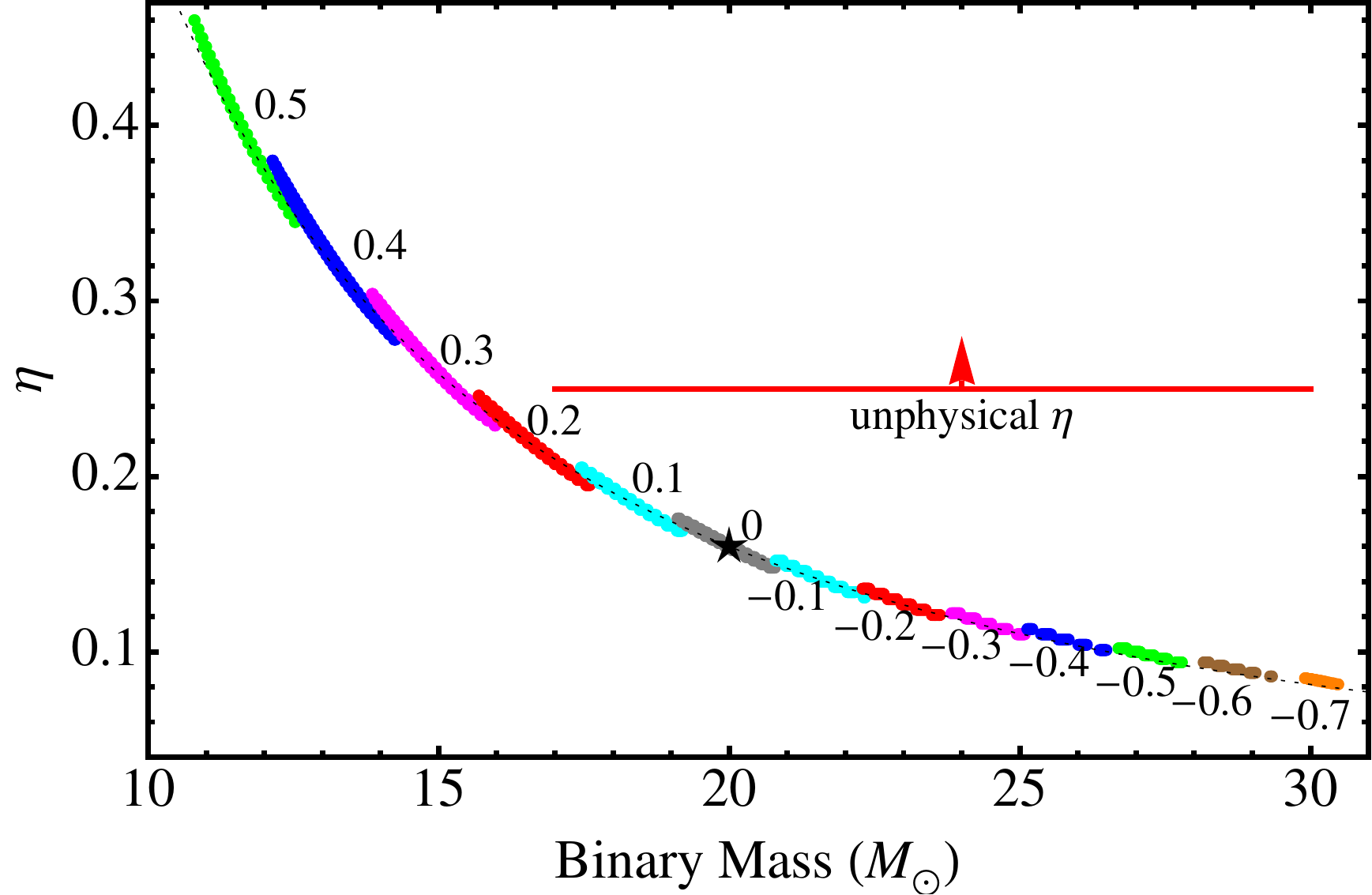}
  \caption{
  \label{fig:M20early} 
  Match between spinning $20 M_{\odot}$ waveforms with non-spinning templates in early aLIGO.  
Each of  the strips shows the region of $M$-$\eta$ space for which the non-spinning waveform 
has a match of  0.97 or higher with a spinning signal.  The total spin is indicated next to each 
region.  For spins  above 0.5 or below -0.7 the best overlap with a non-spinning waveform is 
less than 0.97. The curve of constant chirp mass is indicated by a dotted line.}
\end{center}
\end{figure}

Fig.~\ref{fig:M20early} shows the results for a binary with a total mass of
20$M_\odot$ and mass ratio 1:4 ($\eta = 0.16$), i.e.~a $4M_\odot$--$16M_\odot$
binary, and matches calculated using the early aLIGO spectrum.  We consider
sources with a total spin in
the range [-1,1] in steps of 0.1.  For each configuration, we identify
non-spinning waveforms which give a match of greater than 0.97 with the
spinning source.  When the binary spin is zero, the non-spinning
model matches the target signal at the correct parameters (indicated by a
star), and along a strip in the parameter space with a width of 5\% in mass,
and 10\% in $\eta$.  When the binary contains spinning black holes, the
non-spinning model matches the signal, but with a bias in the mass and mass
ratio. The binary spins are indicated  with different colours, labeled by the
total spin of the binary.  For large-spin signals there are no non-spinning
waveforms that have a match above 0.97. When using the early aLIGO noise
curve, the range of spins for which non-spinning waveforms have a match above
0.97 is $[-0.7,0.5]$.   However, for spins above about $\chi = 0.3$, the
best-match waveform has an unphysical value of the mass ratio $\eta$.  The
curve of constant chirp mass $M \eta^{3/5}$ is indicated by a dotted line. 

\begin{figure}
\begin{center}
  \includegraphics[width=0.5\textwidth]{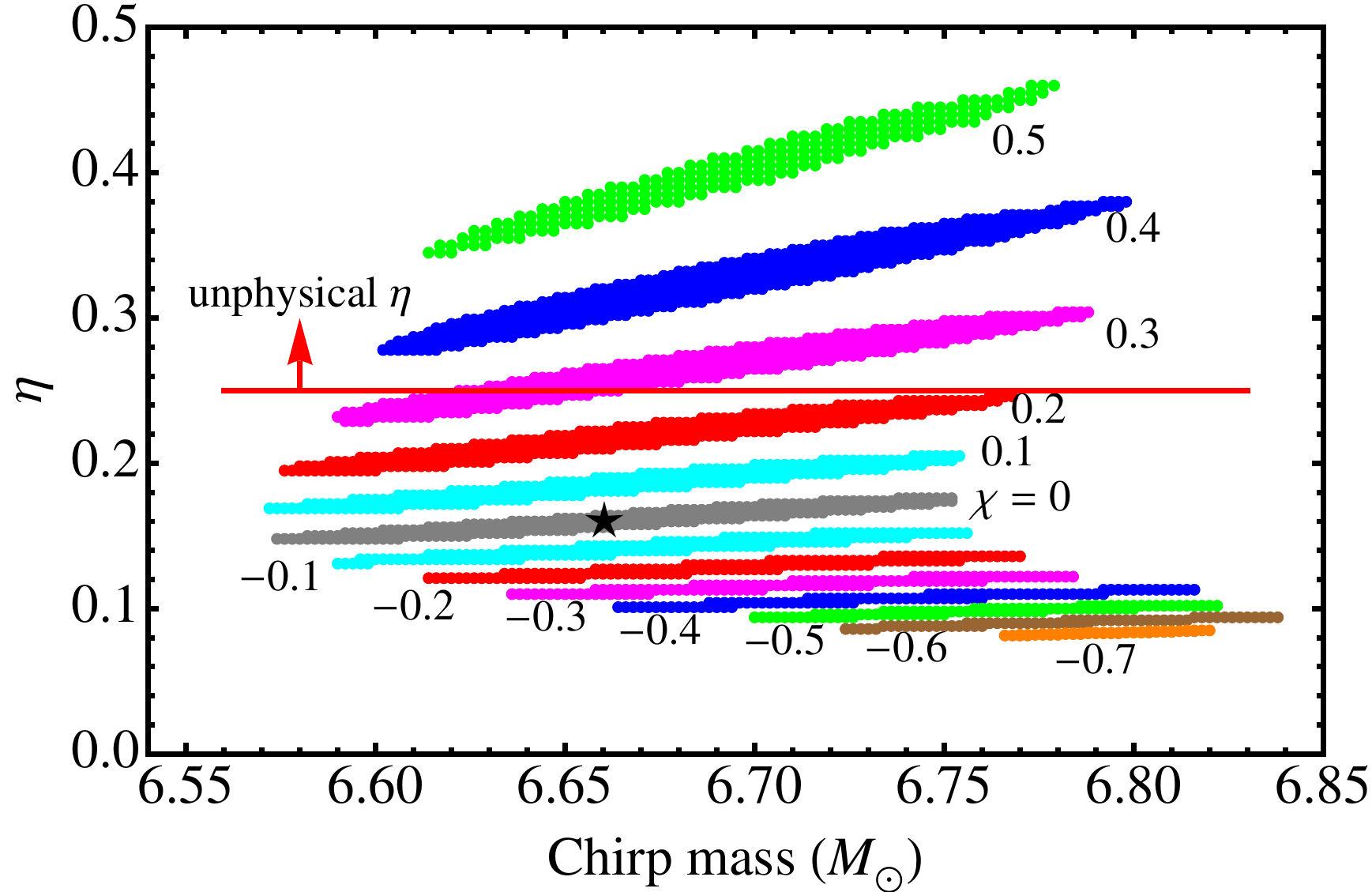}
  \caption{
  \label{fig:M20matchesChirp} 
  Same results as in Fig.~\ref{fig:M20early}, but now parametrized in terms of chirp mass and 
symmetric mass ratio. This figure reinforces the accuracy with which the chirp mass can be 
recovered. 
  }
\end{center}
\end{figure}

The first thing to note is that, in all cases, the chirp
mass of the best matched non-spinning waveform is very similar to the true chirp mass of
the system.  We can see from Fig.~\ref{fig:M20early} that the high-match regions follow a line of 
constant chirp mass. In Fig.~\ref{fig:M20matchesChirp} we
re-parameterize the results of Fig.~\ref{fig:M20early} in terms of $\eta$ and
${\cal M}$. We see that, even though there is a strong degeneracy between $\eta$
and $\chi$, the correct chirp mass ${\cal M} = 6.66 M_{\odot}$ falls within the
high-match region for most values of spin. The deviation from the correct chirp
mass is no more than $\sim3$\% for large anti-aligned spins. 

The second important point relates to the mass-spin degeneracy. 
For aligned-spin binaries ($\chi > 0$), the best match is obtained with a
lower-mass, higher-mass-ratio model, while for anti-aligned spins, the best
matched non-spinning waveform has a higher mass and lower mass-ratio.  This is
to be expected as aligned spins are expected to cause the system to ``hang up''
and chirp more slowly, mimicking a lower-mass signal.  

\begin{figure}
\begin{center}
  \includegraphics[width=0.5\textwidth]{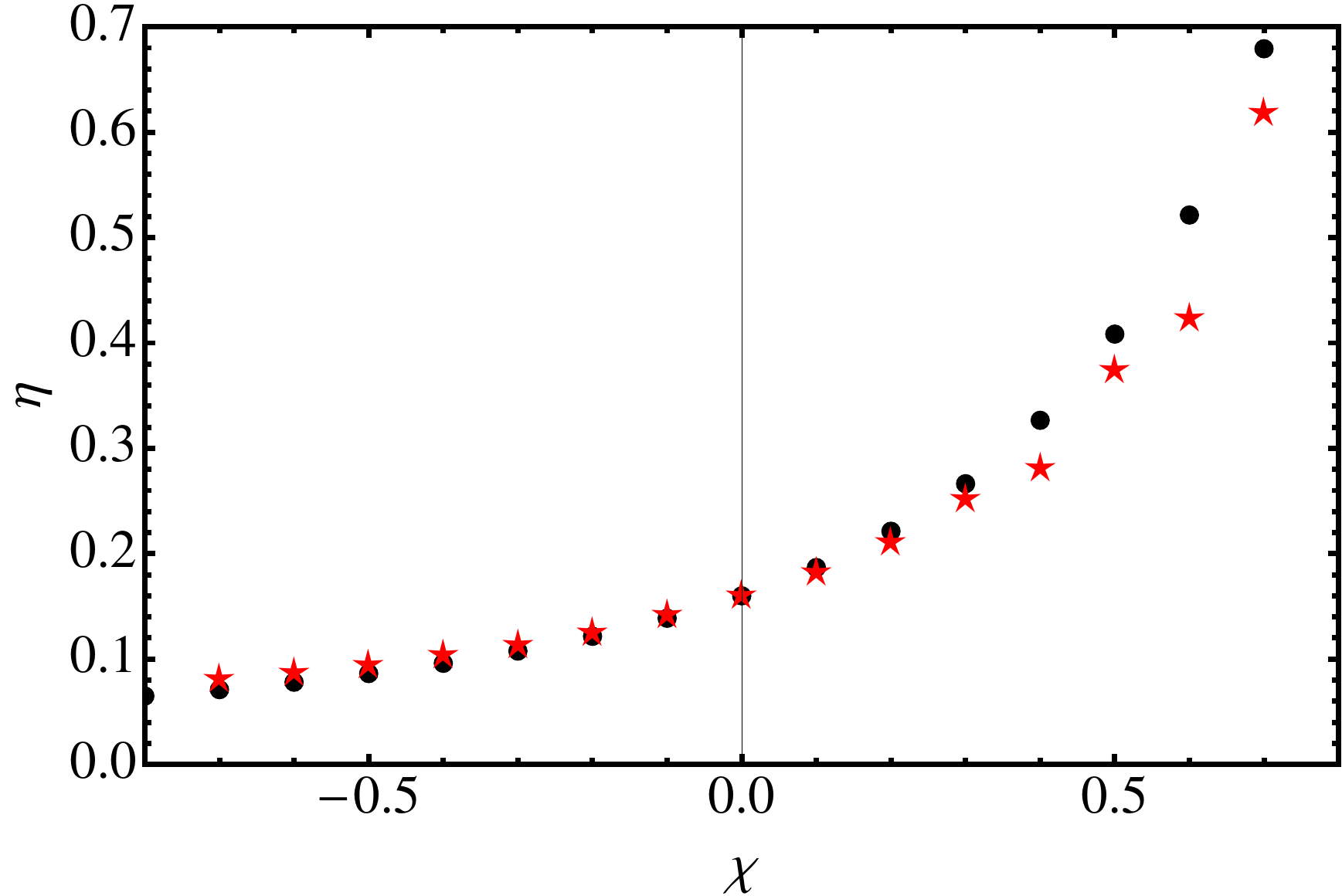}
  \caption{
  \label{fig:degeneracy} 
Comparison of the $\chi$-$\eta$ degeneracy as found for the phenomenological
model (stars), and from the requirement that $\Delta \Psi = \Delta\Psi(\chi_{PN}=0)$ in
Eq.~(\ref{eqn:PNdeg}) for fixed chirp mass, for signals with $M =
20\,M_\odot$ and $\eta = 0.16$. The naive PN prediction agrees well 
with the full IMR results for low spins, and for most anti-aligned spins.  }
\end{center} 
\end{figure}

It is informative to return to the post-Newtonian phasing given in
Eq.~(\ref{eqn:PNdeg}) to see whether the observed degeneracy matches what is
theoretically expected.  In low-mass binaries, the signal is dominated by the
inspiral, which can be represented in a PN expansion, as described in
Sec.~\ref{sec:PNtheory}. The chirp mass is determined to high accuracy by the
leading-order term in the PN expansion of the phase in the Fourier domain, and
we can then solve Eq.~(\ref{eqn:PN}) (with fixed chirp mass) to find the
symmetric mass ratio that mimics the effect of the spin.  We do this by solving
(\ref{eqn:PNdeg}) as a function of frequency (in the detector's sensitive band)
and then averaging over the values of $\eta$ that we obtain.
Fig.~\ref{fig:degeneracy} shows the symmetric mass ratio that corresponds to
each value of the spin, as predicted from Eq.~(\ref{eqn:PN}), and as found in the
mismatch analysis of the phenomenological models that is shown in 
Fig.~\ref{fig:M20early}. We see that the PN estimate is
remarkably close to that found for the full IMR models, at least for the 
$20\,M_\odot$ binaries used in this example. 
 
\begin{figure}
\begin{center}
  \includegraphics[width=0.5\textwidth]{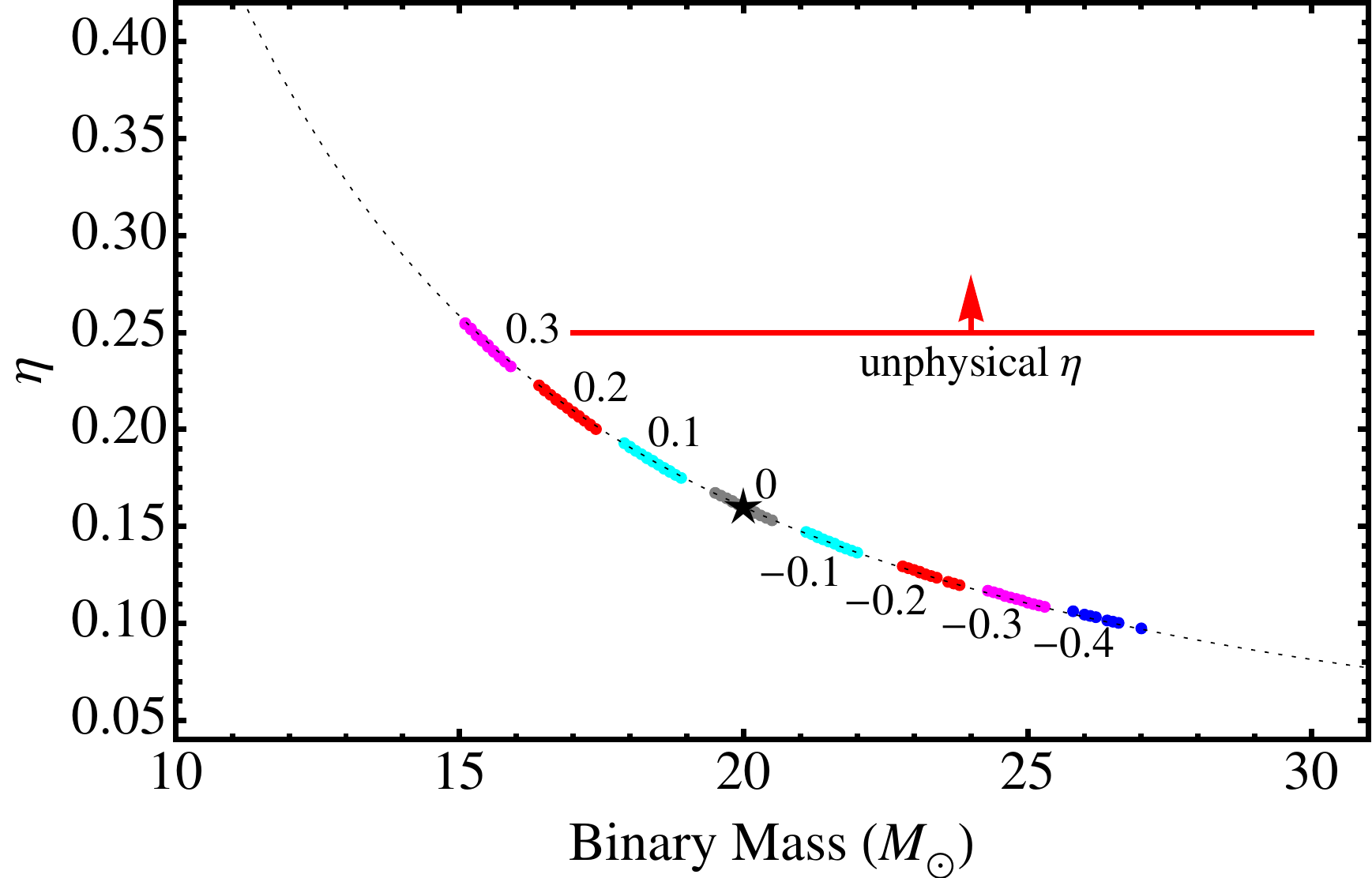}
  \includegraphics[width=0.5\textwidth]{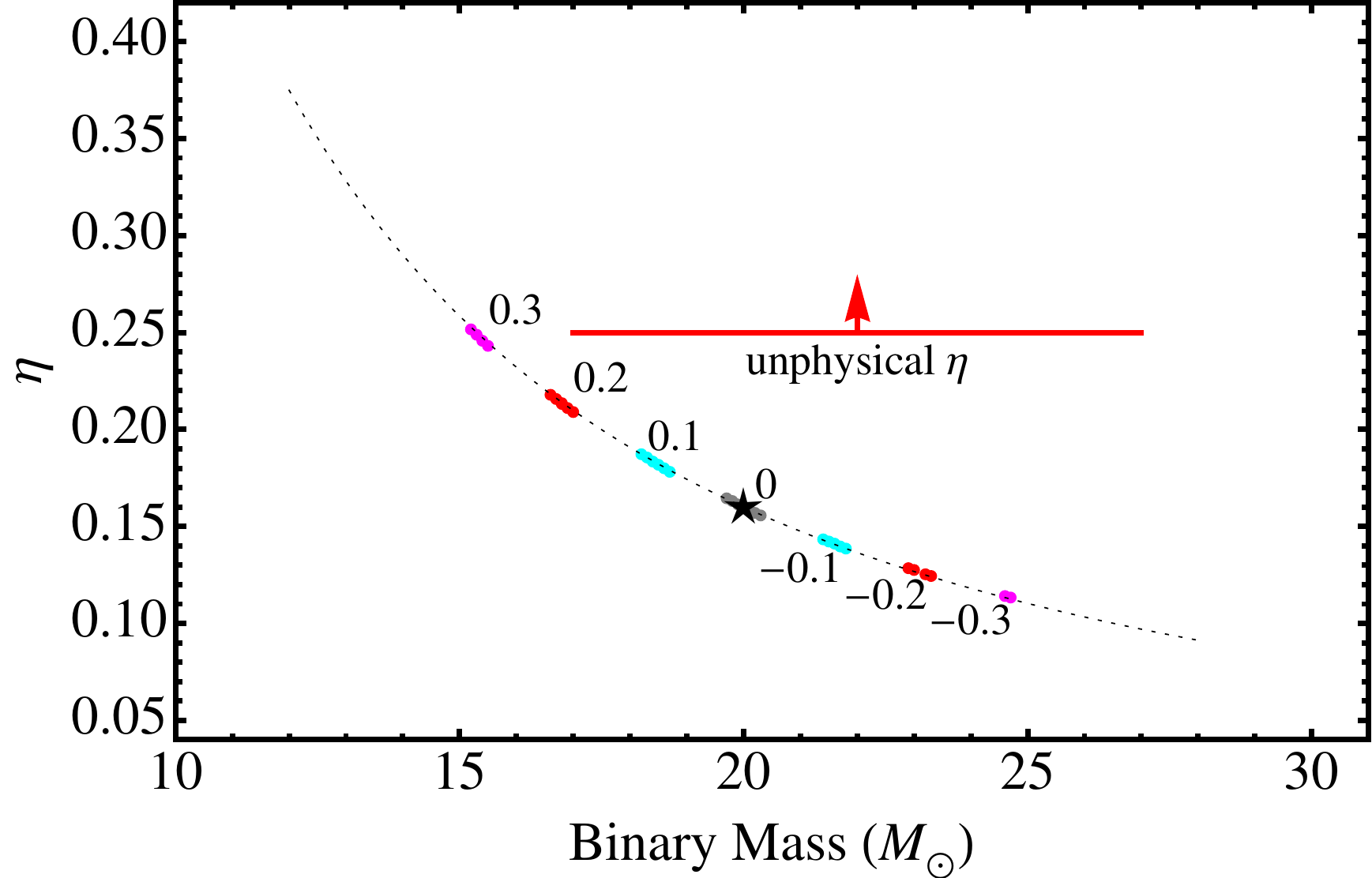}
  \caption{
  \label{fig:M20matches} 
  Matches of spinning $20 M_{\odot}$ waveforms with non-spinning templates in aLIGO at design
  sensitivity.  The upper panel shows the results for the no-SRM configuration, while the bottom 
  shows results for the final sensitivity
}
\end{center}
\end{figure}

\subsection{Evolution of noise curves}

We present results for two design aLIGO configurations in Fig.~\ref{fig:M20matches}: 
the no-signal-recycling (no-SRM) configuration and the final high-power
zero-detuned sensitivity curve (final).  Qualitatively, the results are quite similar
for the no-SRM and final noise curves.  Again spinning signals are recovered
with a match above 0.97 using non-spinning templates, with a decrease in 
recovered mass for positive spin systems.  The chirp mass is still well recovered.  
However, the detector's
extra sensitivity makes it easier to distinguish the spinning signals with a
non-spinning model. The range of spins for which we can find a
non-spinning signal with matches greater than 0.97 is now only $\chi \in [-0.4,0.3]$ for
no-SRM and $\chi \in [-0.3,0.3]$ for the final configuration. We also see that each 
match region shrinks, although
its location is unchanged. In moving from the early noise
curve to no-SRM, the most significant sensitivity improvement is at low
frequency, while the final configuration offers much greater high frequency
sensitivity.  The results for these two curves are quite comparable (and
significantly better than the early curve) indicating that it is low frequency
sensitivity that provides the biggest improvement in breaking the degeneracy.
With the later noise curves, the variation of chirp mass between the spinning
signal and non-spinning template is less than 1\%.
\footnote{The figures may seem to indicate that a $20 M_{\odot}$ binary
with total spin of 0.5 that would be detected by a non-spinning search in the
early aLIGO configuration, but not later ones.  This is of course not true:
the final detector is roughly three times more sensitive at these masses, and
so a signal with SNR 10 in early aLIGO would have an SNR of 30 in the final
configuration.  Even with a match of 0.9, this would still give an SNR of 27,
and the match would be sufficient that it would pass any signal consistency
tests.} 

\section{Mass-spin degeneracy at higher masses}
\label{sec:high_mass}

For black-hole binaries with masses greater than $20 M_{\odot}$ the merger and
ringdown parts of the signal become increasingly important and contribute an
ever increasing fraction of the signal-to-noise ratio.  At higher masses, we
do not expect the chirp mass to determine the waveform to such an extent as
for the $20 M_{\odot}$ system and there is no \emph{a priori} reason to expect
that a degeneracy between mass and spin will persist.   

We begin by considering a $50 M_{\odot}$ binary with mass ratio 1:4.  At this
mass, the merger and ringdown will provide a significant fraction of the
signal to noise ratio.  As a crude estimate of the effect, imagine that the
inspiral part of the waveform is valid up to the innermost stable circular
orbit (ISCO) of a point particle orbiting a Schwarzschild black hole,
which is at 90\,Hz for a $50\,M_{\odot}$ binary.  For the early aLIGO noise
curve, the inclusion of the merger and ringdown will increase the SNR by a
factor of 4.  At the final aLIGO sensitivity, the improved low
frequency response increases the contribution of the inspiral but the merger
and ringdown still contribute a comparable SNR to the inspiral (so that the
overall SNR is  $\sim50$\% higher, because they add in quadrature).  Thus,
there is no reason to expect that the chirp mass is still well recovered or
that the degeneracy discussed previously still holds.

\begin{figure}
\begin{center}
  \includegraphics[width=0.48\textwidth]{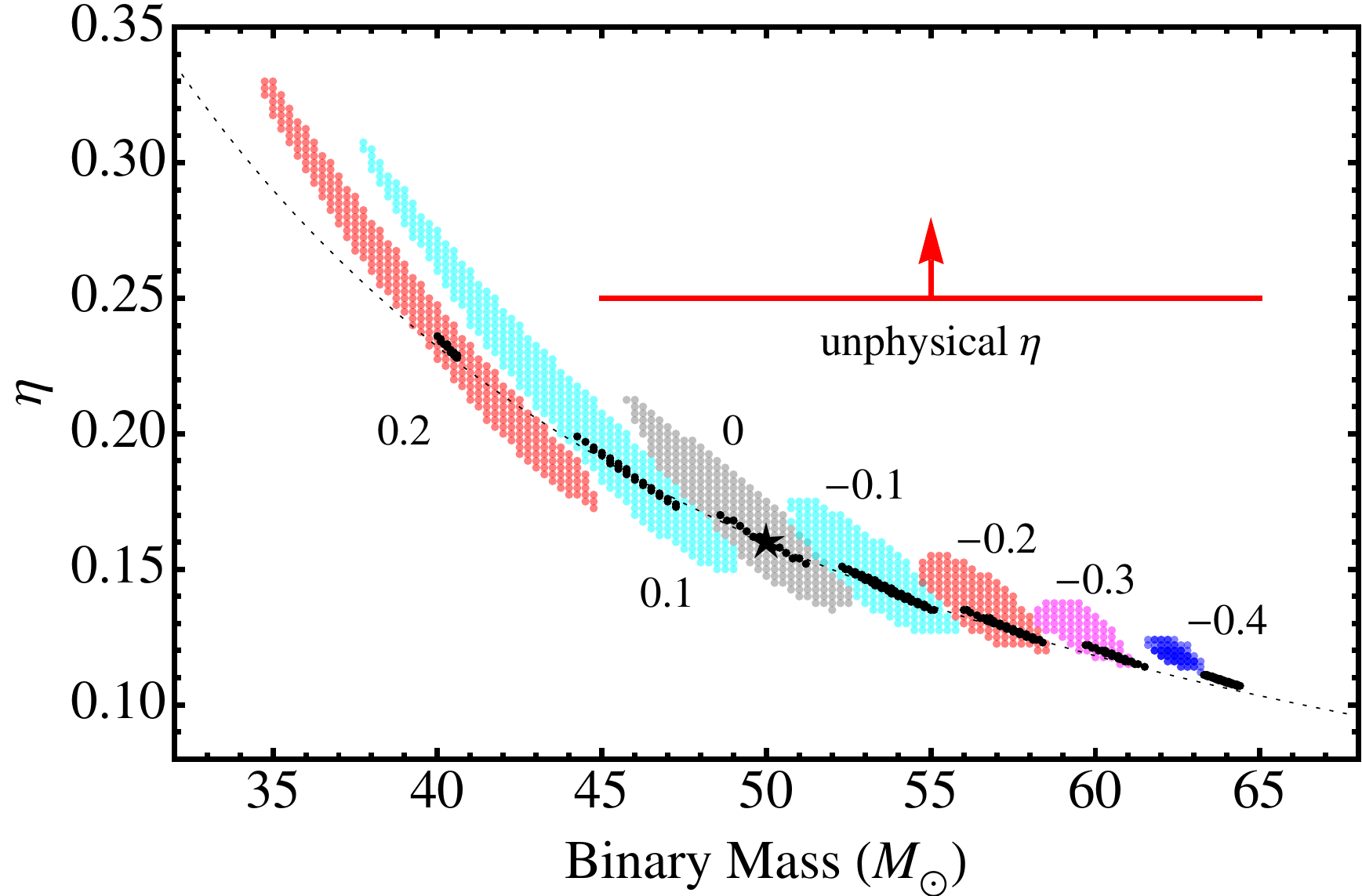}
  \caption{
  \label{fig:M50matches} 
  Matches of spinning $50 M_{\odot}$ waveforms of mass ratio 4:1 with
non-spinning templates in early and final aLIGO.  Each of  the strips shows
the region of $M$-$\eta$ space for which the non-spinning waveform has a match
of  0.97 or higher with a spinning signal.} 
\end{center}
\end{figure}

In Fig.~\ref{fig:M50matches} we again show the regions of the non-spinning
parameter space which give a match greater than 0.97 with a spinning binary
(with values of $ \chi$ from -1 to 1 in steps of 0.1).  Here, we show the results for
both the early and final aLIGO spectra.  The results are qualitatively
similar to the $20 M_{\odot}$ binary with spinning signals well recovered by
the non-spinning waveforms, and the chirp mass accurately recovered.  The
sizes of the regions are roughly consistent with the lower mass system, with
a mass accuracy of $\sim10$\% and range in $\eta$ of 0.05-0.1 for the early
noise curve.  For both early and design curves, signals with spins between 0.2
and -0.4 have matches above 0.97 with non-spinning waveforms.  Interestingly
the degeneracy still roughly follows the line of constant chirp mass, even
though the merger and ringdown contributed significantly to the SNR of the
signal.

\begin{figure}
\begin{center}
  \includegraphics[width=0.48\textwidth]{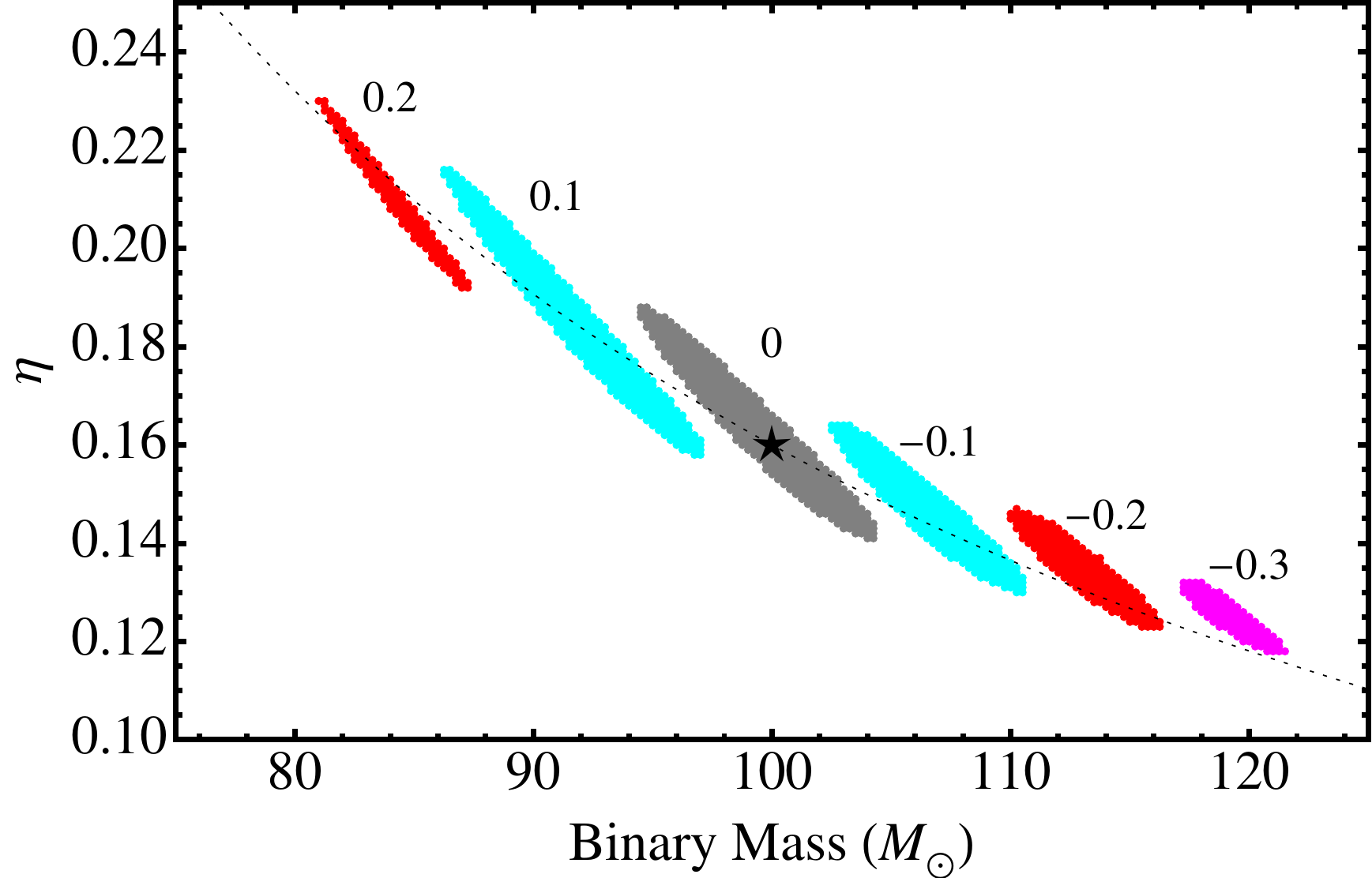}
  \caption{
  \label{fig:M100matches} 
Matches of spinning $100 M_{\odot}$ waveforms of mass ratio 4:1 with
non-spinning templates in design aLIGO.
}
\end{center}
\end{figure}

Next, we increase the mass to $100 M_{\odot}$ and repeat the analysis.  At
this mass, the point-particle ISCO is at 40 Hz, so there is essentially no
power from the inspiral in the initial detectors, and only a small amount in
the early aLIGO.  Even at the final sensitivity, the merger and ringdown provide the vast
majority of the SNR (a factor of 4 more than the inspiral).  Thus one expects
that it is really the merger that dictates the waveform as seen by the
detector.  Figure  \ref{fig:M100matches} shows results for 100$M_\odot$, again
with $\eta = 0.16$.  For aLIGO at final design sensitivity, the curves are again lying 
pretty much along the line of constant chirp mass.

We do not show results for the early aLIGO sensitivity curve in Fig.~\ref{fig:M100matches},
because the results depend strongly on the choice of model, either PhenomB or
PhenomC. The variation of the two models with respect to the physical parameters is
sufficiently large through merger and ringdown that they lead to qualitatively different
results in our mismatch studies, when we use the early aLIGO noise curve, where the 
merger and ringdown contribute essentially all of the SNR. The two models were 
developed as pioneering models of aligned-spin IMR waveforms for use in 
searches, and for this we expect that they are sufficient; but their fidelity with respect
to parameter estimation at high masses has not yet been tested, 
and should be a focus of future work to refine them.

\section{Parameter recovery}
\label{sec:parameters}

Once a black hole merger has been detected, we wish to extract the signal 
parameters as accurately as possible.  Parameter estimation proceeds by identifying
regions of parameter space which give a signal that is most consistent with the data.  It
stands to reason that these regions will contain waveforms that have a good match with 
the observed signal.  Thus we might expect that confidence regions in
parameter estimation are associated with regions of high match between signal and template.
In this section, we show this expectation to hold in detail in the high-SNR limit and derive an
expression for the value of the match that corresponds to a given confidence at a given SNR.

We begin by presenting the argument, using the Fisher matrix formulation,
and then we use this connection to re-interpret our earlier results in terms of parameter recovery.
We follow the formalism used Ref.~\cite{2008PhRvD..77d2001V}, but provide only a
brief discussion of the Fisher matrix formalism, and refer the reader to the
Ref.~\cite{2008PhRvD..77d2001V} (and references therein) for more details.

\subsection{Connection between mismatch and confidence regions}

Let us assume that a signal $h_{0}$ is present in the data.  
The detector data is given as
\begin{equation}\label{eq:data}
  s(t) = h_{0}(t) + n(t) \, .
\end{equation}
To investigate signal recovery and parameter extraction at leading order, we Taylor 
expand the signal in a region of the true parameters ($\theta = 0$) as 
\begin{equation}\label{eq:wf_expansion}
  h(\theta) = h_{0} + \theta^{i} h_{i} + \ldots
\end{equation}
Here $h_i = \partial_{i} h$ is used to denote the derivative of the waveform
with respect to the paramters $\theta_{i}$, which will include $M$, $\eta$,
$\chi$ as well as the amplitude, phase and coalescence time that are maximized
over in the match calculations.  

The likelihood, for a given set of $\theta_{i}$,  is 
\begin{equation}
  p(s| \theta) \propto \exp \left\{- \frac{(s - h(\theta) | s - h(\theta) )}{2} \right\} \, .
\end{equation}
Substituting the expressions for $s$ (\ref{eq:data}) and $h(\theta)$ 
(\ref{eq:wf_expansion}) into the above and keeping leading order terms gives
\begin{equation}\label{eq:like}
  p(s| \theta) \propto \exp \left\{ - \frac{(n|n)}{2} + \theta_{i} (n|h_{i}) 
  - \frac{\theta_{i} \theta_{j} (h_i | h_j)}{2} \right\}
\end{equation}  

In the context of Bayesian parameter estimation, this can be recast in terms of a
posterior probability distribution for the parameters $\theta_{i}$ using Bayes theorem:
\begin{equation}
  p(\theta | s) \propto p(s | \theta) p(\theta)
\end{equation}
where $p(\theta)$ is the prior probability distribution for the parameters.
In what follows, we use a uniform prior on the parameters.  In general, such a prior
will not be physically motivated but, for a detectable signal, the likelihood will be
peaked in a small enough region of parameter space to make this approximation
reasonable.  

Given the above, we are interested in calculating the expected offset between the true 
parameters and the mean value from the posterior distribution.  We also want to evaluate the
size of a confidence region containing a given fraction $p$ of the posterior probability.  These
quantities give us two different measures of the expected accuracy of parameter recovery.

We begin by calculating the mean of the parameter $\theta_{i}$ as
\begin{equation}\label{eq:mean}
  \langle \theta_{i} \rangle = \int d\theta \theta_{i} \, p(\theta | s) 
  = (h_{i} | h_{j} )^{-1} (n | h_{j}) \, .
\end{equation}
Thus, the mean of the posterior distribution will be offset from the true
parameter values due to the presence of noise.  One way to characterize this is the
expected size of the error waveform, $h_{E} = h(\langle \theta_{i} \rangle) - h_{0}$ as
\begin{equation}
  \langle h_{E}^{2} \rangle_{n} =  \langle (n | h_{i}) (h_{i} | h_{j} ) (h_{j} | n) \rangle_{n}
  = k \, ,
\end{equation}
where $k$ is the dimension of the parameter space and $\langle \rangle_{n}$ 
indicates the expectation value over many noise realizations.  Thus, on average
the difference between the true signal and ``best fit'' waveform will have an
amplitude of $\sqrt{k}$.

Next, we turn our attention to confidence regions in parameter space  --- 
a region $\Theta$ of parameter
space that contains a given probability $p$ of the posterior distribution,
\begin{equation}
  p = \int_{\Theta} d\theta p(\theta | s) \, .
\end{equation}
There
are many ways to construct such a region, and one typically also requires the 
smallest possible region.  To calculate confidence regions in the Fisher 
approximation, we
begin by observing that the covariance between parameters is given as
\begin{equation}\label{eq:covariance}
  \langle \theta_{i} \theta_{j} \rangle = (h_{i} | h_{j} )^{-1} \, .
\end{equation}
Using this expression and Eq.~(\ref{eq:mean}), we can re-express the posterior
distribution as
\begin{eqnarray}\label{eq:posterior}
  p(\theta | s) &\propto& \exp \left\{- \frac{1}{2} (\theta_{i} - \langle \theta_{i} \rangle) (h_{i} | h_{j})
  (\theta_{j} - \langle \theta_{j} \rangle) \right\} \nonumber \\
  &\simeq& \exp \left\{- \frac{1}{2} (  | h(\theta) - h(\langle \theta \rangle) |^{2} ) \right\} \, .
\end{eqnarray}
Then, the minimum volume region which contains a fraction $p$ of the posterior
probability is the one for which 
\begin{equation}\label{eq:conf}
  | h(\theta) - h(\langle \theta \rangle) |^{2}  < \chi^{2}_{k}(1-p) \, ,
\end{equation}
where $\chi^{2}_{k}(1-p)$ is the chi-square value for which there is $1-p$ probability
of obtaining that value or larger, and $k$ denotes the degrees of freedom, determined by
the number parameters included in $\theta_{i}$.  At leading order, the confidence 
interval contains all points for which the amplitude of the difference between the model 
and best fit waveforms lies below the given threshold.

There are six parameters in the aligned-spin waveform model: $M$,
$\eta$, $\chi$, $A$, $\phi_{c}$ and $t_{c}$.  When calculating matches, we have
maximized over the latter three parameters (amplitude, phase and time) and 
reported match over the three dimensional space of mass, mass ratio and spin.  
Thus, we have calculated the three dimensional matches $M(h_1 | h_2)$, 
maximized over A, $\phi_c$ and $t_c$.  It is straightforward to re-cast our earlier results
in terms of mismatches.  To do this, we note that
\begin{equation}\label{eq:diff_as_match}
  \min_{A_2} | h_{1} - h_{2} |^{2} = 
  |h_{1}|^{2} \left\{1 - \frac{(h_{1} | h_{2})^{2}}{|h_{1}|^2 |h_{2}|^{2}}\right\} \, .
\end{equation}
When we restrict to the subspace of masses and spins, we can write the final term as the
match between the waveforms.

By substituting Eq.~(\ref{eq:diff_as_match}) into Eq.~(\ref{eq:conf}), we see that the 
confidence region is defined by all points in the parameter space for which the match satisfies
\begin{equation}\label{eq:match_thresh}
  M(h(\theta), h(\langle \theta \rangle) \ge 1 - \frac{\chi^{2}_{k}(1-p)}{2 \rho^2}
\end{equation}
where the value of $k$ is given by the dimension of the remaining parameter space.
This expression gives us a straightforward way to re-interpret the match calculations
presented earlier.  For example, with a three dimensional parameter space, the 90\%
confidence region at a given SNR is given by:
\begin{equation}
  M(h(\theta), h(\langle \theta \rangle) \ge 1 - \frac{3.12}{\rho^2}
\end{equation}
which has the nice property that a match of 0.97 corresponds to a 90\% confidence 
region at an SNR of 10.  For a two dimensional parameter space, the right hand side is 
$1 - 2.3/\rho^2$, meaning that a match of 0.97 corresponds to 90\% confidence region
at an SNR of 9.

We should note that there are 
numerous assumptions used in the derivation of these results.  Most significantly,
all of the Fisher matrix results consider only leading order effects and become less
reliable at lower SNR, see e.g., Ref~\cite{2008PhRvD..77d2001V} for a detailed discussion 
of the issues.  When actually calculating confidence regions for a signal, detailed
parameter estimation analyses calculate the posterior distribution (\ref{eq:posterior}) and 
integrate to find the region containing 90\% of the probability.
Here, we are using a hybrid approach: we use the result of the Fisher matrix calculation
to decide the threshold on match required to define the given confidence region, but then 
calculate the match between waveforms exactly, without recourse to any approximations.
Furthermore, we are maximizing the match over three dimensions, analytically for $A$ and 
$\phi$ and using a Fourier transform to search over time.  Thus, we are only applying the
Fisher result to the three dimensional subspace of mass, mass ratio and spin.  Even there, we
are merely using the calculation to determine the appropriate match threshold: the matches
are calculated exactly.   Consequently, the match regions should be in good agreement with
the 90\% confidence regions.  We verified that for an SNR of 20, the region identified by our
match criteria did contain 90\% of the probability to within $\pm0.5\%$. 

\subsection{Implications for detectability and parameter estimation} 

Let us now return to the results of Sec.~\ref{sec:spin_eta} and \ref{sec:high_mass}
and use the relationship derived above to re-interpret the results in terms of
confidence intervals.  We can interpret Fig.~\ref{fig:M20early} in two ways: in the context
of either a non-spinning or spin-aligned templated search.  If we were to perform a
search with non-spinning templates and observe a $M = 20 M_\odot$, $\eta = 0.16$ 
binary with non-spinning components with 
SNR 9, then the 90\% confidence region would lie along a strip in the parameter space with 
a width of 5\% in mass, and 10\% in $\eta$.  For a binary containing the same mass black holes 
with spins of -0.5, the 90\% confidence region would be of approximately the same size 
(5\% in mass, and 10\% in $\eta$) but centred at $M = 27 M_\odot$, $\eta = 0.10$.%
\footnote{Strictly speaking, if the best match between the signal and a model waveform
is less than unity, the relation between match and confidence region in equation (\ref{eq:match_thresh})
must be modified to reflect this.  The 90\% confidence region will contain all points with a
match of 0.97 or greater with the \textit{best fit} model waveform.}
The same holds for other values of the spins: the reported statistical uncertainty in 
parameters is relatively small, while the systematic errors can be significant.

Alternatively, we can consider the three dimensional space of $M$-$\eta$-$\chi$.  In 
that case, a match above 0.97 corresponds to a 90\% confidence region at SNR
of 10.  Thus, for example, Fig.~\ref{fig:M20early} 
shows that the 90\% confidence region for a signal with $m_{1} = 16, m_{2} = 4$ 
and $\chi \in [-0.7, 0.5]$ would contain waveforms with non-spinning components.
Figures \ref{fig:M20matchesChirp} to \ref{fig:M100matches} can be interpreted in a
similar manner.  
For higher SNRs, the confidence intervals shrink.  At SNR 10 they correspond to 
matches of 0.97, SNR 20 to 0.992 and SNR 30 to 0.9965.

\begin{figure}
\begin{center}
  \includegraphics[width=0.48\textwidth]{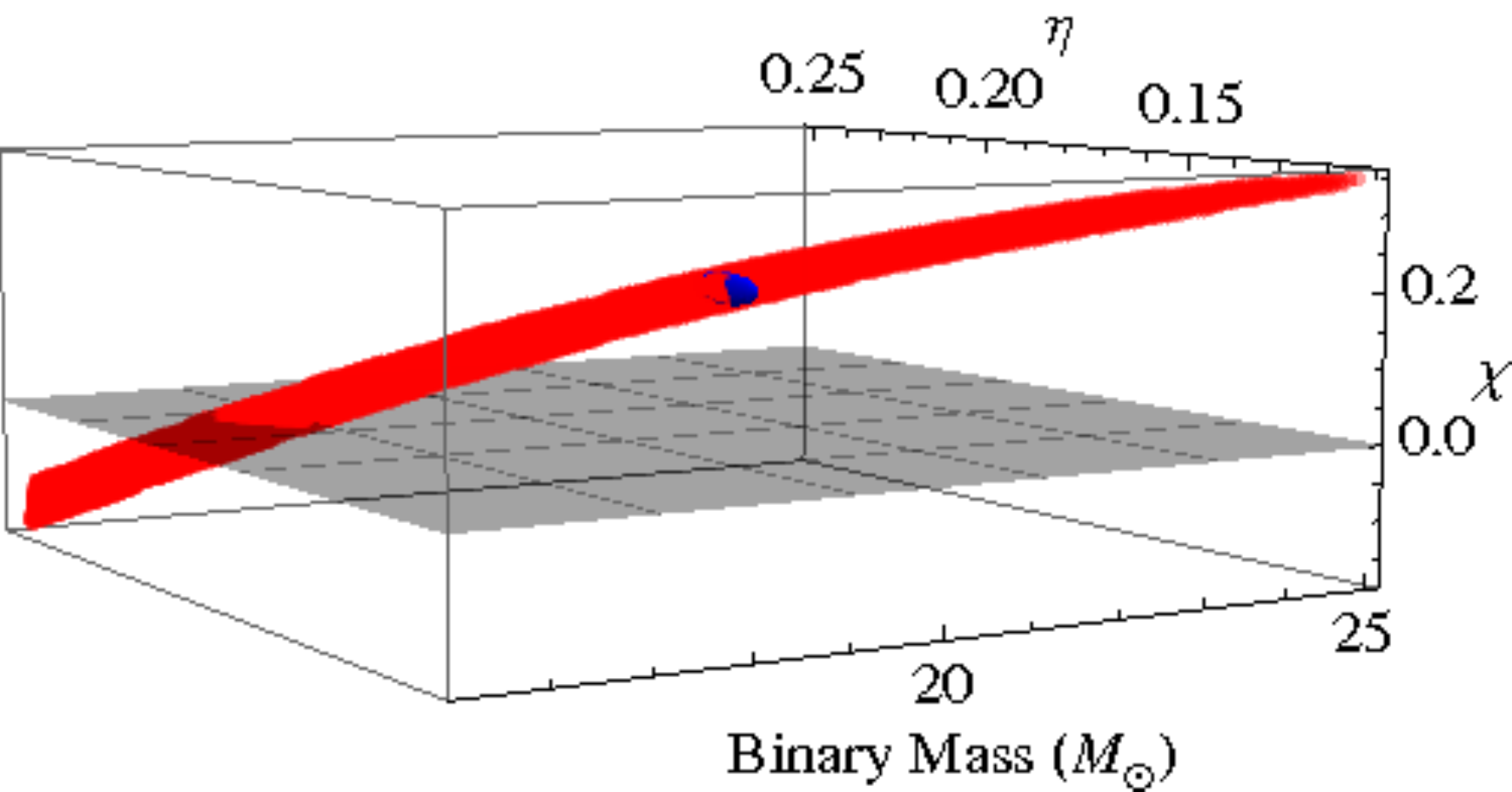}
   \includegraphics[width=0.48\textwidth]{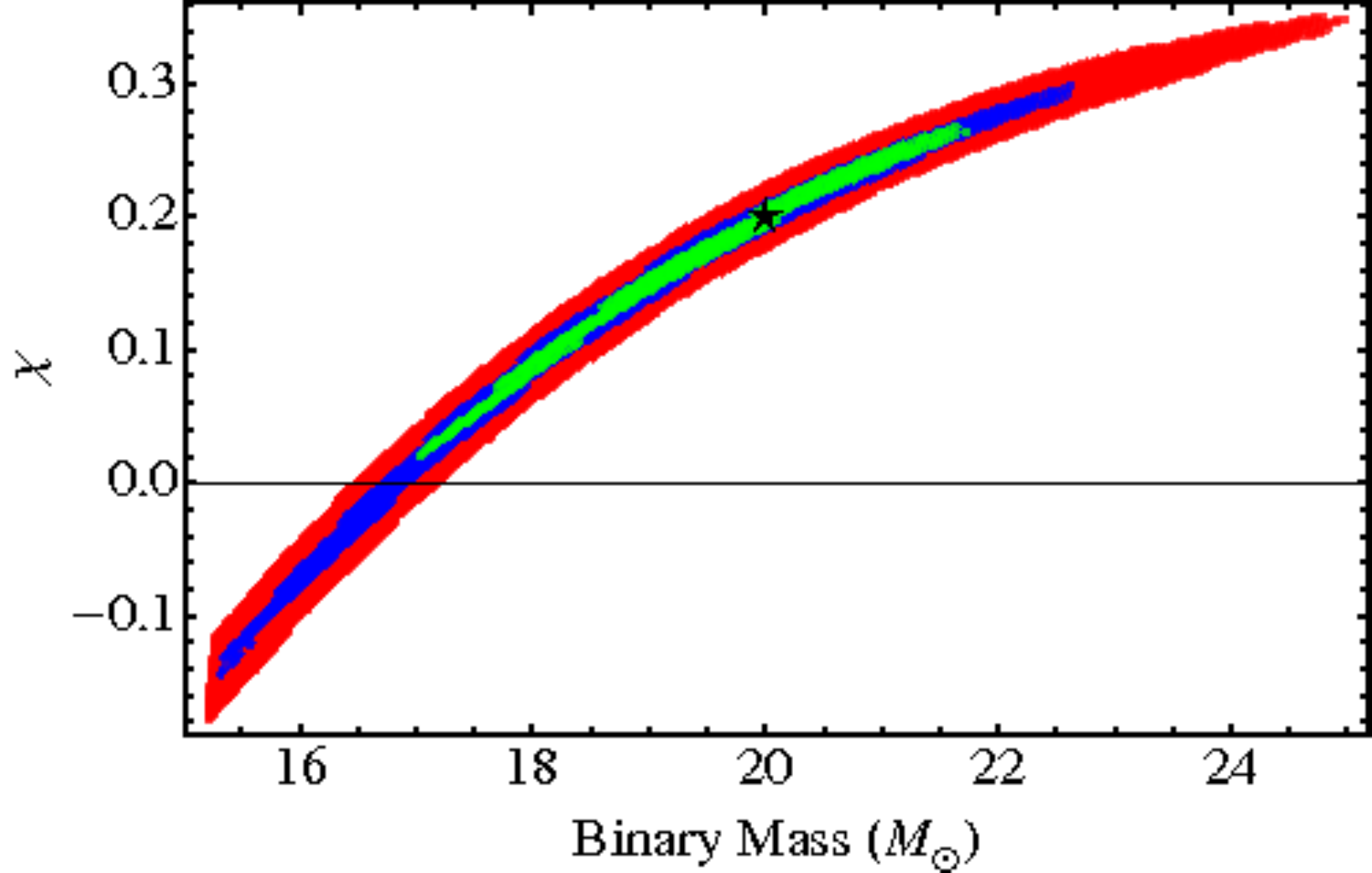}
    \includegraphics[width=0.48\textwidth]{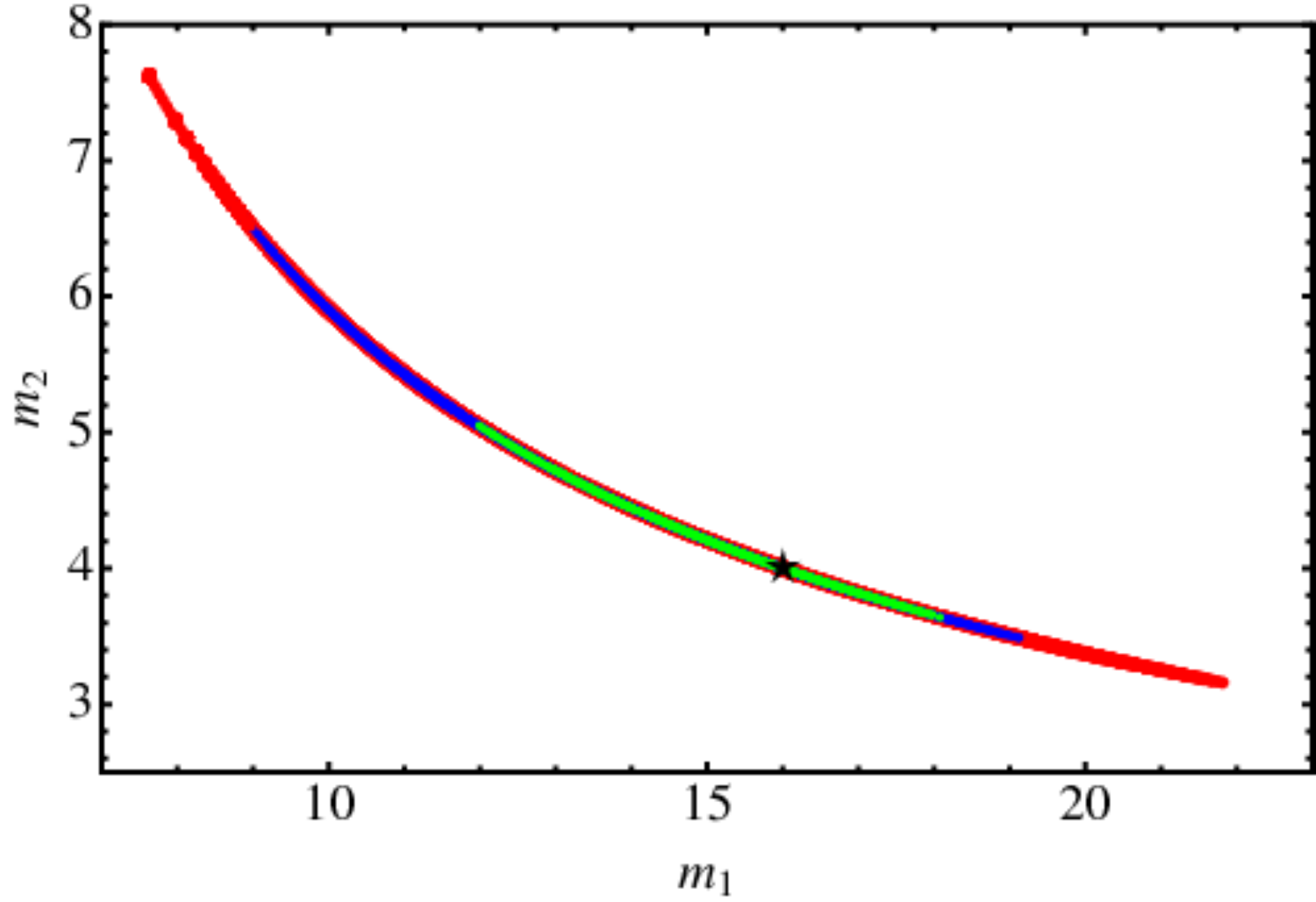}
  \caption{
  \label{fig:3Dconfidence} 
90\% confidence intervals for a 20\,$M_\odot$, 1:4 ($\eta = 0.16$) signal with 
spin $\chi = 0.2$, using the design aLIGO noise curve.
The top panel shows the full three-dimensional confidence region for SNR=10; 
the gray surface indicates the $\chi = 0$ plane, and corresponds to the $\chi = 0.2$ region in 
Fig.~\ref{fig:M20matches}. The lower two panels show the same data projected 
onto the $M$-$\chi$ and $m_1$-$m_2$ planes, and also indicate the 90\% confidence
regions for SNRs 20 (blue) and 30 (green).
The true physical parameters are indicated by a ball (top panel), or star (lower
panels). Regions with unphysical $\eta$ ($>0.25$) are not shown.
  }
\end{center}
\end{figure}

\begin{figure}
\begin{center}
  \includegraphics[width=0.48\textwidth]{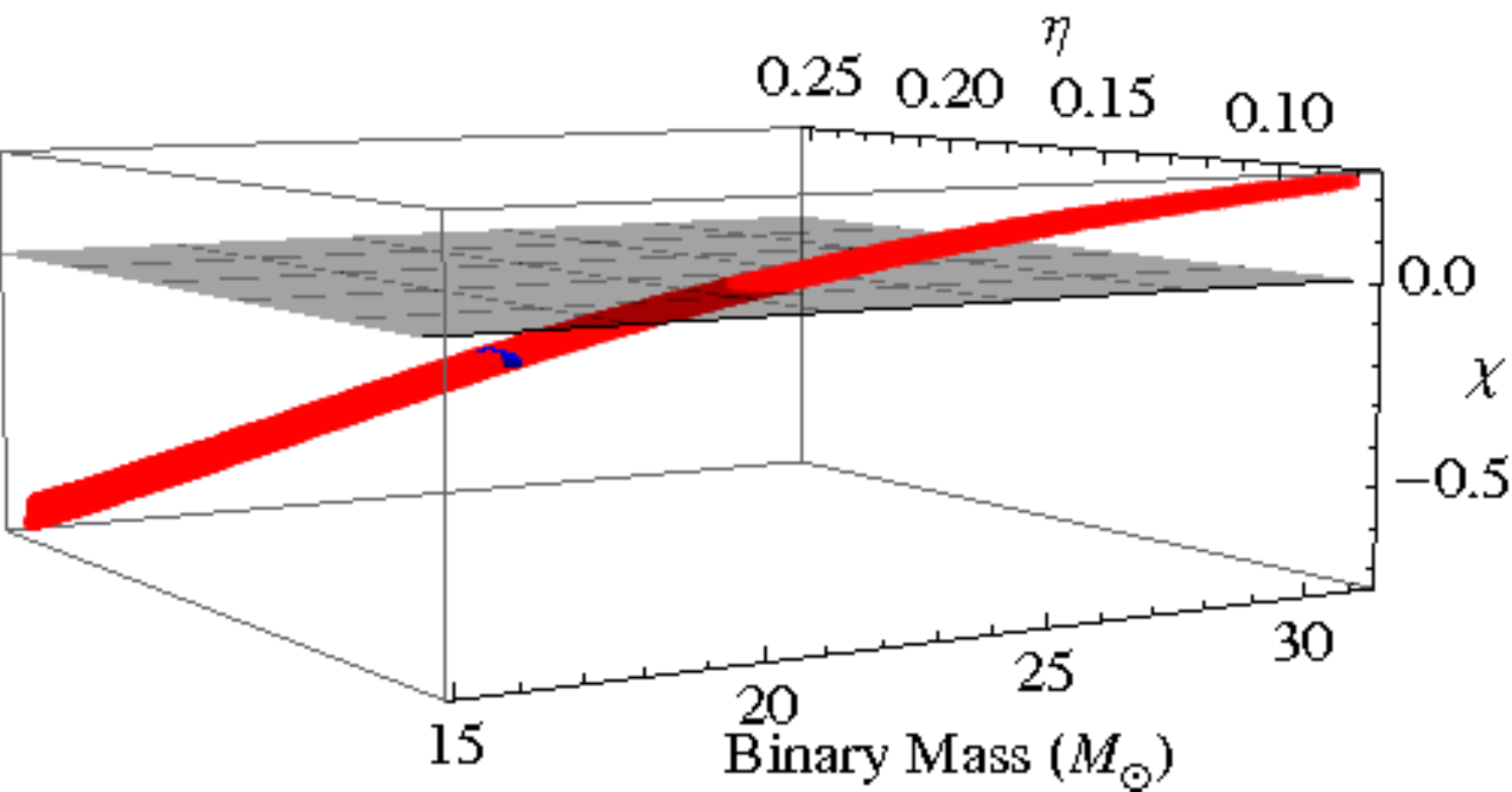}
   \includegraphics[width=0.48\textwidth]{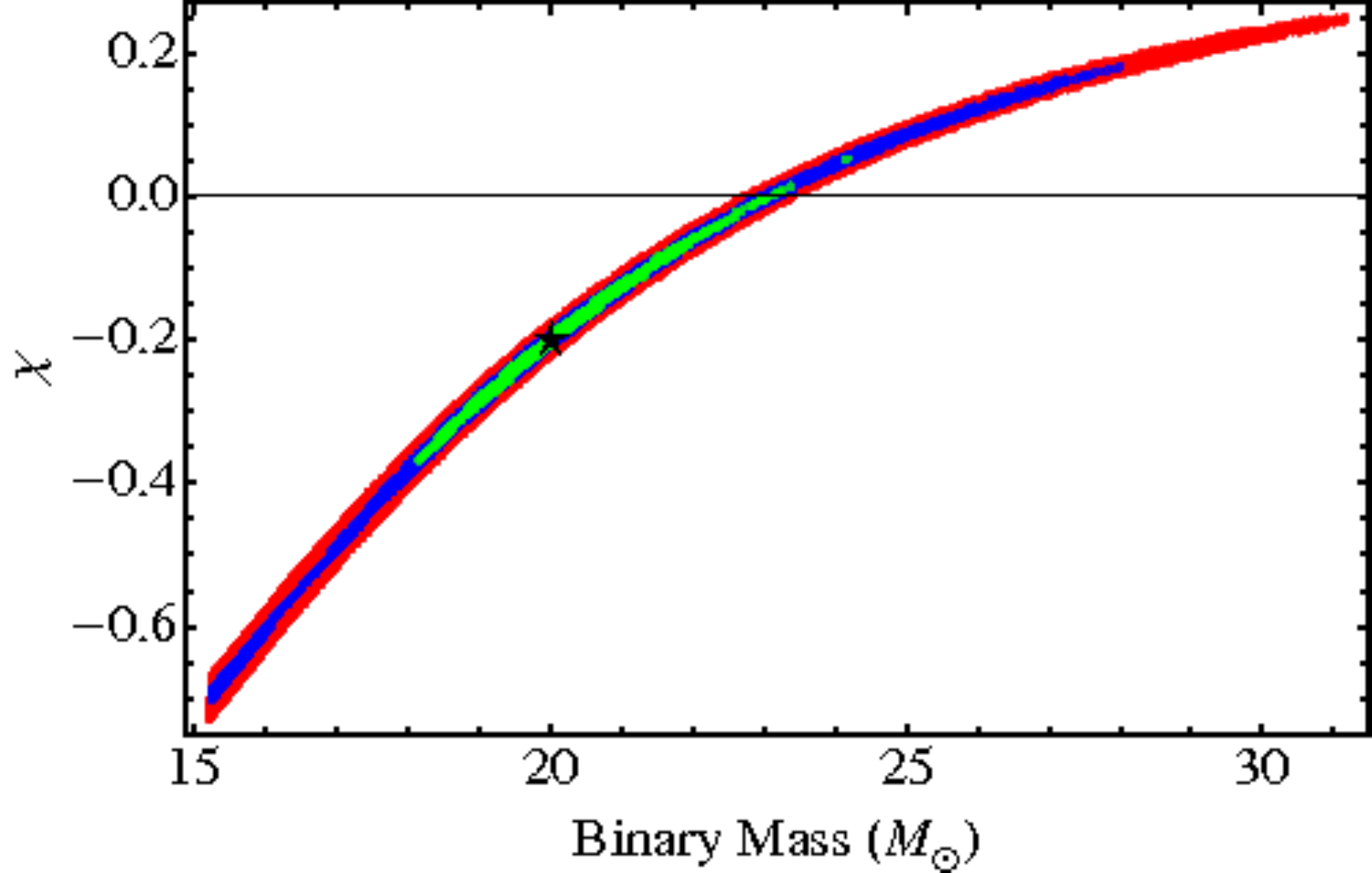}
    \includegraphics[width=0.48\textwidth]{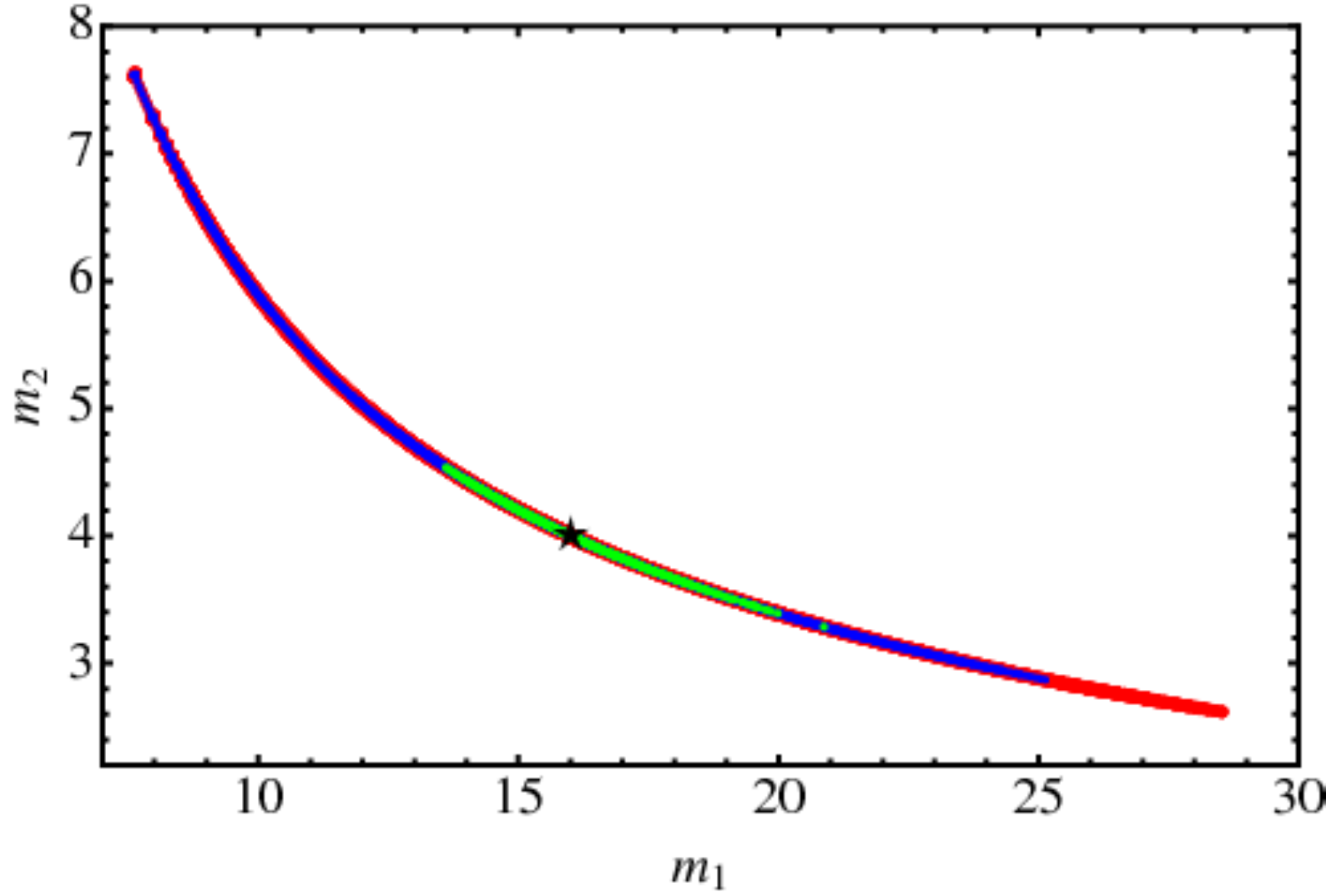}
  \caption{
  \label{fig:3DconfidenceB} 
Same as Fig.~\ref{fig:3DconfidenceB}, but now the signal has spin $\chi = -0.2$.
  }
\end{center}
\end{figure}

We can use the analysis of the previous section to estimate the 90\% confidence intervals if we 
employ an aligned-spin model for parameter estimation. For signals with an SNR of 10, the
three-dimensional confidence region in $(M,\eta,\chi)$ will correspond to matches greater
than 0.97. In these cases we calculate the matches between a given aligned-spin signal, and
all other aligned-spin model waveforms with varying mass, mass-ratio and total spin. 
In Fig.~\ref{fig:3Dconfidence} we show the 0.97 match volume
for a 20\,$M_\odot$ 1:4 binary with $\chi = 0.2$, using the final aLIGO noise curve. The
top panel shows the full three-dimensional confidence region, and in the lower panels the
results are projections onto the $M$-$\chi$ and $m_1$-$m_2$ planes, to aid the interpretation.  
Figure \ref{fig:3DconfidenceB} shows similar results, but for a binary with $\chi = -0.2$.

Since our waveform model now includes
spin, the best match is unity at the correct parameters. But from the figures we see that the 
90\% confidence region extends well beyond the correct parameters, and is far from the 
naive image conjured by the term ``error ellipse''. We note in particular that this volume 
includes a region of the non-spinning binary parameter space; the intersection with the 
$\chi = 0$ plane is consistent with the lower panel of Fig~\ref{fig:M20matches}. 
If we were to estimate the
parameters of this signal with an aligned-spin waveform model, we would find that the best-fit 
parameters indicated that the black holes were spinning, but could not rule out that they might
in fact be non-spinning --- or indeed have spins with the opposite orientation. 

The reason for this large uncertainty in the parameters is that the SNR is low; it is only
10. For an SNR of 20, the 90\% confidence interval corresponds to a match of above
0.992, and for an SNR of 30, the match must be above 0.9965. We see that for these
higher SNRs, the confidence region shrinks. However, only at SNR 30 is a non-spinning
signal excluded from the 90\% confidence region for the $\chi = 0.2$ case, and even at this 
SNR (which may be quite rare in aLIGO observations), the smaller mass is determined to within only 
25\%. When $\chi = -0.2$ (Fig.~\ref{fig:3DconfidenceB}, the 90\% confidence region includes 
spins as high as $\chi = -0.7$, even when the SNR is 20. Note also that in these figures we have 
included only those portions of the confidence regions with $\eta \leq 0.25$, i.e., physically 
acceptable values; the confidence
region would extend further in the upper two panels if the model were not constrained.

\section{Conclusions}
\label{sec:conclusion}

We have investigated the degeneracy between mass-ratio and spin in gravitational 
waveforms, going beyond inspiral models to include merger and ringdown signals from 
black-hole-binary mergers. We have used phenomenological inspiral-merger-ringdown
(IMR) models to study the subset of the full binary parameter space that includes 
non-spinning black holes, and black holes with spins parallel or anti-parallel to the orbital
angular momentum of the binary. 

A degeneracy between mass-ratio and spin is already known for inspiral signals, which are
relevant to ground-based gravitational-wave detectors for masses $M \leq 12$\,$M_\odot$ 
\cite{Buonanno:2009zt,Brown:2012qf}.
We find that this degeneracy persists at higher masses, where the detectors are also sensitive to 
the merger and ringdown. This means that in a GW search that uses only non-spinning binary 
templates (and this is computationally cheaper than including spin), the signal may still
be detected, but the best-match template will have strongly biassed parameters. 
The mass-ratio--spin degeneracy follows lines
of constant chirp mass, so the chirp mass will be recovered with 
reasonable accuracy, even up to high masses and for spinning signals. However, the total mass 
and mass-ratio of the best-matched template will be biased. 
As shown in Fig.~\ref{fig:M20early}, binaries with higher aligned spins will be recovered with a 
higher mass ratio, while those with anti-aligned spins will be recovered with a lower mass ratio. 
If we restrict the search to physical values of the symmetric mass ratio $\eta$, then 
comparable-mass binaries whose spins are aligned with the orbital angular momentum
will tend to be missed in the search. We show how these results will
change as the detector evolves towards its final sensitivity in Fig.~\ref{fig:M20matches}, and 
for higher masses in Figs.~\ref{fig:M50matches} and \ref{fig:M100matches}.

We also demonstrated that it is possible to use match calculations to estimate the confidence 
intervals in parameter estimation. For example, the 90\% confidence region for the three-dimensional
parameter space of ($M$, $\eta$, $\chi$) for signals with SNR 10 is given by the region with 
matches above $\sim0.97$. This allows us to estimate how the mass-rato--spin degeneracy
will be reflected in parameter estimation. We show that for modest SNRs ($\sim10$) it may 
be difficult to determine whether a binary includes spin, and even for high SNRs ($\sim30$), 
the mass-ratio--spin degeneracy impairs the accurate recovery in the individual black-hole
masses; see Figs.~\ref{fig:3Dconfidence} and \ref{fig:3DconfidenceB}. These results will affect
the astrophysical conclusions that can be drawn from GW observations, and this will be explored
in more detail in future work. It would also be interesting to exploit our knowledge of the
mass-ratio--spin degeneracy in a jump proposal for Bayesian parameter-estimation 
codes~\cite{Sluys:2008a, Sluys:2008b, Veitch:2010}.

Our results are restricted to aligned-spin binaries, and will be affected by the inclusion of 
precession effects.  At present, no complete inspiral-merger-ringdown model for precessing
systems exists.  It seems unlikely, however, that the inclusion of precession effects will 
reduce the size of the confidence regions. In general it is far more likely that increasing the 
dimensionality of the model parameter space will increase the parameter uncertainty, 
because the confidence regions will now correspond to lower matches.
The recent results in Ref.~\cite{Schmidt:2012rh} show that precession effects have only a
weak impact on the phasing, which suggests that while the inclusion of precession 
is unlikely to break the degeneracy discussed here, it is also unlikely to 
introduce an \emph{additional} degeneracy in these three parameters. 
On the other hand, the inclusion of higher harmonics, which were ignored in this study, may 
improve the accuracy of the parameters. The net impact of these two effects remains to be 
studied in more detail in future work. 

\section*{Acknowledgments}

S.F. and M. H. were supported by Science and Technology Facilities Council grants 
ST/H008438/1 and ST/I001085/1. S.F. also acknowledges the support of the Royal Society.
We are particularly grateful to P. Ajith, Sascha Husa and Frank Ohme for sharing with us their 
implementations of the PhenomB and PhenomC waveform
families and match-calculation code. 
We also thank Chris Messenger and John Veitch for useful 
discussions.  %

\bibliography{refs}

\end{document}